\documentclass[reprint,onecolumn,notitlepage,showpacs,preprintnumbers,amsmath,amssymb,aps,pra]{article}
\usepackage[pdftex]{graphicx}
\usepackage{amsmath}
\usepackage{amssymb}
\usepackage{amsthm}
\usepackage{graphicx}
\usepackage{url}
\usepackage{dcolumn}
\usepackage{bm}
\usepackage{psfrag}
\usepackage{footnpag}
\usepackage{nccfoots}
\usepackage{stmaryrd}
\usepackage{color}
\usepackage{authblk}
\usepackage{sidecap}
\usepackage{mathtools}
\usepackage{caption}
\usepackage{float}

\begin{document}

\title{Compact-form solution to the time-dependent Schr\"odinger equation with an arbitrary potential}
\author{Ivan Gonoskov \thanks{ivan.gonoskov@gmail.com\,,\;ivan.gonoskov@uni-jena.de}}
\affil[]{Institute for Physical Chemistry, Friedrich-Schiller-University Jena, Max-Wien-Platz 1, 07743 Jena, Germany}
\date{\today}

\maketitle

\begin{abstract}
We obtain exact solutions to the class of parabolic partial differential equations of arbitrary dimensionality and with arbitrary potentials. The solutions are presented in a compact-form: as explicit mathematical expressions consisting of finite number of standard mathematical operations with finite (condition-independent) number of variables in a discrete or continous region of consideration. The general approach to obtain the compact-form solutions combines a number of methods in the fields of partial differential equations, enumerative combinatorics, and number theory, which we describe in detail. We discuss their advantages and perspectives for the various fields in physics, mathematics, and advanced calculus.
\end{abstract}

\newpage
\section*{Preliminaries}\,

The central idea of our manuscript is to figure out the simplicity-complexity of the expressions for the solutions to the differential equations and the ways one can obtain one in a possibly most suitable form. Thus, before we consider our main topic related to the solutions of certain partial differential equation, let us discuss the central concept of our study: "compact-form solutions". There are several well known types of mathematical expressions which differ by the abstract complexity and convenience. These types are typically discussed in terms of "satisfactory" or "elegant" forms from the viewpoint of the efficiency of counting the considered expressions, see for example \cite{StanleyComb}. 

Assuming the great advantage of explicit expressions (over implicit) at least for the analytical study, we can range the mentioned types as follows. The simplest and most desired form is a closed-form expression. This expression consists of a number of simplest accepted mathematical operations and functions and does not include limits, derivatives, integrals(sums) and special functions. The analysis of such expressions is almost straitforward. Slightly more complex form is analytical expressions, which typically can include in addition some integrals (of known accepted functions) and some known special functions. Lastly, the general explicit expressions can include in addition limits, derivatives, and (multiple) integrals with multiple (unlimited number) variables.

In the current study we propose to define an additional type: "compact-form expressions" which can be placed on the complexity-siplicity line between the general explicit and the analytical expressions, see the diagram below.\\  
\begin{center}
  \includegraphics[scale=0.37]{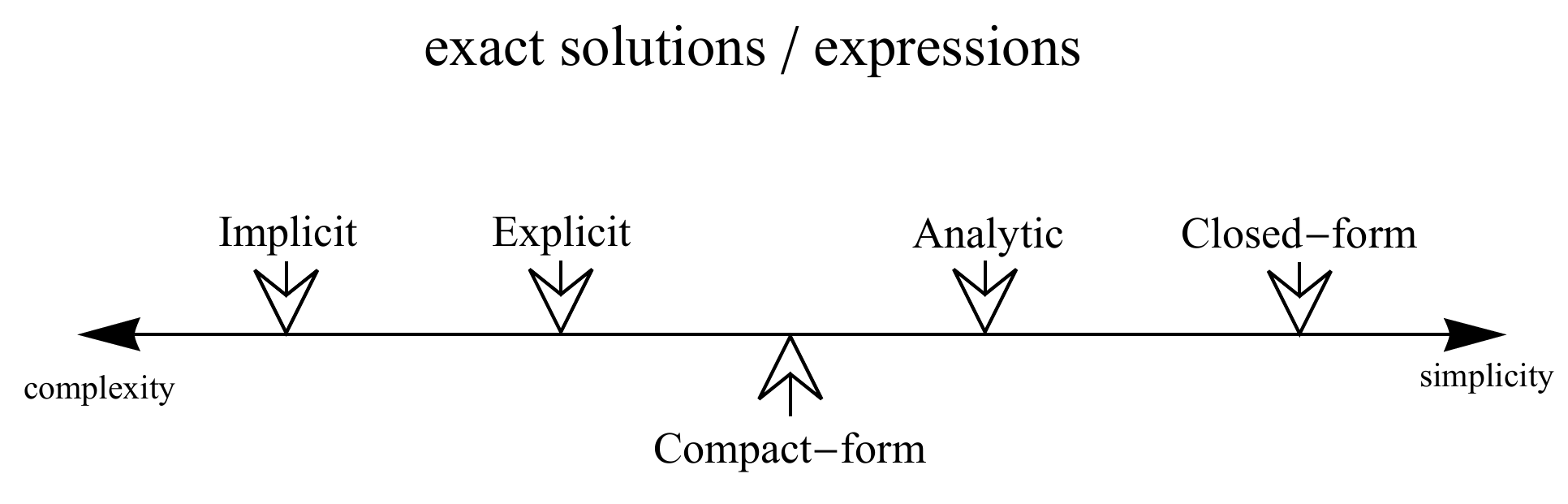}
\end{center}
The main difference of compact-form from the analytical form is a possible presence of limits and integrals of arbitrary accepted functions (which is not of crucial difficulty for the present computational tools). The crucial difference from the explicit form is that it includes only finite number of variables (some constant independent from any problem variables/conditions) and therefore only limited number of integrals/sums. It is assumed also that compact-form expressions exclude continued recursive procedures like Euclidian division as can be related to reformulation of multi-index notations, see also \cite{3term}. Now, let us formally define the compact-form solution and consider some related examples.
\newpage
{\bf Definition:} a compact-form solution is a mathematical expression which is an exact explicit solution of the corresponding partial differential equation in a discrete or continuous region and constists of a finite discretization independent number of accepted mathematical operations with a finite discretization independent number of associated variables in the whole expression.

{\textit{Note:}} under the "accepted mathematical operations" we assume the following operations: "$+$","$-$","$\div$","$\,\cdot\,$", sums, products, integrals, of the combinations of functions like $f_{p}(x)=x^{\alpha}$ ($\alpha\in\mathbb{R}$), $f_{e}(x)=\exp(x)$, $f_{ln}(x)=\ln(x)$ in the real or complex domain which are smooth enough for the considered problem. We assume that the potentials or other internal functions in the partial differential equations and the initial states constist of a combination of accepted mathematical operations and functions as well. 

{\textit{Proposal:}} a number of the unique-title variables in the expression is called a variable-weight and denoted as $W_{expr}$. (Nonessential renaming does not count).
 
Let us now consider a simple example to clarify the described concept. Assume we have a differential equation for the function $f(x)$ of one variable $x$ with a condition $f(0)=1$. There is an internal arbitray given function $u(x)$ which is continuous and integrable in the region of consideration. The differential equation reads:
\begin{equation}\label{i01}
\frac{df}{dx}=u(x)\cdot{}f\;\;\nonumber
\end{equation}

The exact solution of this equation as a compact-form expression can be obtained straightforwardly: $f(x)=\exp\bigg[ \int\limits_{0}^{x} u(y) dy \bigg]$. Since the function $u(y)$ is arbitrary, the given solution with the related integral can be formally not an analytic expression. The variable-weight of this solution is $W=2$.

On the other side, one of the usual ways to obtain general explicit solutions of differential equations (usually more complicated) leads to the following evident expression to the considered equation, see also \cite{COD} and references therein:

\begin{equation}\label{i02}
f(x)=1 + \int\limits_{0}^{x} u(x_{1})\, dx_{1} + \int\limits_{0}^{x} u(x_{1}) \int\limits_{0}^{x_{1}} u(x_{2})\, dx_{2}\, dx_{1} + \ldots \;.\nonumber
\end{equation} 

This form of solution is not a compact-form expression since its variable-weight $W=\infty$. Even when a discrete domain for $x$ ($N$ intervals) is considered for an approximate solution with an error proportional to $N^{-p}\; (p\in\mathbb{N})$ \cite{SSP}, the variable-weight of such expressions is typically $W\sim{}N$. Thus, this recursive type solution is not a compact-form expression.

The evident advantage of a compact-form expression over the general explicit one is that the former gives a straightforward access to all the sum/integral-terms which allows one to use various tools of advanced calculus and analysis. This can be, for example, renumbering, substitutions, using method of differences and changing of variables, finding asymptotics, and many others.

\newpage
\section*{Introduction and Statement}

The fundamental question we address in this manuscript is about the useful and convenient for the analisys form of the solution to the time-dependent Schr\"odinger equation (TDSE). Here we consider an initial value problem for the most general multi-dimensional equation with an arbitrary given time-dependent potentials, and also some similar partial differential equations (PDE). We start from the one-dimensional case of TDSE, the formulation is following:
\begin{align}\label{i1}
i\,\frac{\partial}{\partial{}t}\,&\Psi=\left[-\frac{1}{2}\,\frac{\partial^{2}}{\partial{}x^{2}}\,+U(x,t)\right]\Psi\,,\\
&\Psi(x,t)|_{t=0}=\Psi_{0}(x)\,.\nonumber
\end{align}

For some specific potentials $U(x,t)$ (quadratic potential, for example) the solution to this problem $\Psi(x,t)$ is known in closed-form. There are also several known potentials for which the exact solutions in analytic form were derived. However, in general, the exact solution of the problem Eq.(\ref{i1}) is known for the arbitrary smoth $U(x,t)$ in a special form of Neumann series with successive integration. Usually this form of the solution is referred as Dyson series \cite{Dyson}, or modified Born series \cite{Born}, or Cyclic Operator Decomposition \cite{COD}. For the Hamiltonian operator $\hat{H}(t)=\left[-\frac{1}{2}\,\frac{\partial^{2}}{\partial{}x^{2}}\,+U(x,t)\right]$ of Eq.(\ref{i1}) the solution reads:

\begin{equation}\label{i2}
\Psi=\left[1+(-i)\int\limits_{0}^{t}\hat{H}(t')\,dt'+(-i)^{2}\int\limits_{0}^{t}\hat{H}(t')\int\limits_{0}^{t'}\hat{H}(t'')\,dt''\,dt'+\ldots\right]\Psi_{0}\,.
\end{equation}
Following this way, one needs to consider an unlimited number of variabeles ($t,\,t',\,t'',\,...$) associated to the sequence of integrals for the analysis of the system evolution in a finite interval $[0,t]$. The variable-weight of this form of solution is $W_{\text{Eq.(2)}}=\infty$. Even for a certain prescribed accuracy level and guaranteed convergence of related series, the estimation of the number of required variables is a separate problem for the each certain $\hat{H}(t)$.

Sometimes it is possible or even very reasonable for solving the Eq.(\ref{i1}) to make a discretization using the time-slicing approach. This means that the considered time interval $[0,t]$ is divided into a number ($N$) of smaller intervals. Implying that the overall error of such solving procedure is proportional to $N^{-p}\; (p\in\mathbb{N})$ this could lead to the useful accurate discrete reformulation of Eq.(\ref{i2}) with the variable-weight $W_{\text{N-discret:\,Eq.(2)}}\gtrsim{N}$, see also \cite{SSP}.

One another special form of solution of problem Eq.(\ref{i1}) is usually related to Feynman's paths integral interpretation of quantum mechanics \cite{Feynman}. The method uses a Lagrangian approach to construct a path integral as a limit of a sequence of integrals from a special slicing
approximation. Under certain assumptions it has been proven that Feynman’s original definition of a path integral gives the fundamental solution to the Schrödinger equation. The related solution can be written as an expression including a path integral of the form $\int_{R}\,F\;dx_{N}\ldots{}\,dx_{i}\,\ldots\,dx_{1}$,
where $R$ is some region of consideration related to a physical interval, $F$ is a function depending on action, and $x_{i}$ are integration variables related to the time-space slicing approximation and forming the piecewise classical path. As in the previous example Eq.(\ref{i2}) the distinguishing feature of this expression is an unlimited ($N$-dependent, $N\rightarrow\infty$) repetative integration with certain order using an unlimited number of different integration variables. 

In some cases, the discrete analog of the problem Eq.(\ref{i1}) is actually finding the solutions of some recurrence relation with number dependent coefficients. This corresponds to a variety of linear multistep methods originated from the famous forward Euler method for solving the ordinary differential equations. In case of three-term recurrence relation with arbitrary number-dependent coefficients the solution can be obtained using the recursive sums with problem-dependent number of indexes and can be further transformed by using the extra-range variables and the floor function for the integer division (see \cite{3term} for the details). Thus, if we limit ourselves here to only the simplest mathematical operations (exclude the repetitive integer division and other operations not listed in compact-form definition), then all the above discussed solutions in this section have the variable-weight of either $W_{\text{c}}=\infty$ (continuous) or $W_{\text{d}}\gtrsim{N}$ (discrete), i.e. are not compact-form expressions.

As one of the main results of the current manuscript, we have obtained the compact-form solution of Eq.(\ref{i1}). The solution for arbitrary time $T$ reads:
\begin{equation}\label{i3}
\psi \left( x,T\right) = \lim_{N\rightarrow\infty}\,\Bigg[\, \sum ^{3^{N}-1}_{M=0}\,P_{M}\cdot \Psi _{0}\Big(\, x\,+\,S_{M}\cdot \tau\, \Big) \,\Bigg]\,, 
\end{equation}
where $P_{M}$ is a TEG-path numbered with $M$-index, $S_{M}$ is an accumulative shift of the path-$M$, $\tau=\sqrt{T/2N}$, and $N\in\mathbb{N}$ is a time-slicing number. The compact-form expressions for the $P_{M}$ and $S_{M}$ are derived using CSI method, all the details can be found in the sections below. We show the typical (one of many options) compact-form expression for the solution:\\
\begin{equation}\label{i4}
\resizebox{.9\linewidth}{!}{
\begin{math}
\begin{aligned}
&P_{M}\cdot \Psi _{0}\Big(\, x\,+\,S_{M}\cdot \tau\, \Big) \,=\,\dfrac{1}{4\pi^{2} }\,E_{M}(x,T,N)\times\,\sum\limits_{k=-N}^{N} \\
&\int\limits ^{\pi }_{-\pi }\int\limits ^{\pi }_{-\pi }  \prod ^{N-1}_{n=0}\left[ (1-2\mathrm{i})+\mathrm{i}e^{-\mathrm{i}\gamma}e^{\mathrm{i}3^{n}\lambda}+\mathrm{i}e^{\mathrm{i}\gamma}e^{2\mathrm{i}3^{n}\lambda}\right]\cdot\Psi _{0}(x+k\tau) e^{\mathrm{i}k\gamma}e^{-\mathrm{i}M\lambda }d\lambda d\gamma \,,\\
&E_{M}(x,T,N)= \\
&\exp\bigg[\dfrac{-\mathrm{i}T}{4\pi^{2}N }\sum\limits_{\kappa=0}^{N}\sum\limits_{s=-N}^{N}\int\limits ^{\pi }_{-\pi }\int\limits ^{\pi }_{-\pi } G_{\small \textcircled{\small \raisebox{-.9pt} {3}}} \cdot U(x+s\tau,T-\frac{{\kappa}T}{N}) e^{\mathrm{i}s\varphi-\mathrm{i}(2^{\kappa+1}-1)\phi}d\phi d\varphi \bigg],\\
&G_{\small \textcircled{\small \raisebox{-.9pt} {3}}}=G_{\small \textcircled{\small \raisebox{-.9pt} {3}}}\left( M , \phi , \varphi \right)\,=\,\dfrac{1}{2\pi}\times\\
&\int\limits ^{\pi }_{-\pi}\prod ^{N-1}_{n=0}\left[(1+e^{\mathrm{i}2^{n}\phi})+(1+e^{\mathrm{i}2^{n}\phi - \mathrm{i}\varphi})e^{\mathrm{i}3^{n}\theta}+(1+e^{\mathrm{i}2^{n}\phi + \mathrm{i}\varphi})e^{2\mathrm{i}3^{n}\theta}\right]e^{-\mathrm{i}M\theta }d\theta.
\end{aligned}
\end{math}
}
\end{equation}

The exact compact-form solution Eqs.(\ref{i3}-\ref{i4}) has a variable-weight $W_{\text{}}=13$. The compact-form expression Eq.(\ref{i4}) includes the exponential part of $M$-th TEG-path $E_{M}(x,T,N)$ which depends on arbitrary given time-depemdent potential $U(x,T)$. Among a number of technical mathematical operations a special place has a (variety-order 3) path-selecting function $G_{\small \textcircled{\small \raisebox{-.9pt} {3}}}(M, \phi , \varphi)$. This function has been obtained using the CSI method with the base representation uniqueness, which is discussed in detail below.

Although we do not discuss at the current stage the further transformations and possible simplifications of Eqs.(\ref{i3}-\ref{i4}), we believe that the considered methods may be useful for a variety of areas and problems, not limited by partial differential equations of parabolic type. Among others, Translational Evolution Grid (TEG) approach allows to include different physical interactions on different (discretized) space-time grids. The Combination-Selection Integration (CSI) method allows to account each unique object or subset of objects from a set, and provides options of advanced calculus for the further analysis. It is naturally useful for the problems of enumerative combinatorics, counting paths on a grid, and analysis of problems with sequential noncommuting transformations.

\section*{Time-slicing and single-step propagators}

In this section we start to obtain a discrete analog of the solution of Eq.(\ref{i1}) that is approximate solution with a precisely calculated accuracy depending on a grid. Further, the discrete solution can be transformed into a compact-form and can be optionally used to obtain a continuous compact-form solution. In particular we consider a time-slicing approach and a single-step propagator technique for the solving of Eq.(\ref{i1}). For the solution at an arbitrary time $T$ consider the following partition:
\begin{equation}\label{ts1}
0<t_{1}<t_{2}<\,\ldots\,<t_{N}\,=\,T\;,
\end{equation}
where $N\in\mathbb{N}$ is a time-slicing number. Now the idea of a construction of the approximate solution is following: since according to the Eq.(\ref{i2}) a small time-interval solution is known, one can recursively construct the multi-step solution using the solution at a previous step $(n-1)$ as an initial state for the step $n$. The considered procedure can be formally written as follows:
\begin{equation}\label{ts2}
\Psi_{n}(t_{n})=\hat{P}_{n}\,\Psi_{n-1}(t_{n-1})-R_{n}\;,
\end{equation}
where $\hat{P}_{n}$ is a single-step propagator (it depends on $\hat{H}(t)$ when solving Eq.(\ref{i1})), and $R_{n}$ is a step error which is typically proportional to some degree of the time-step $\Delta_{n}=t_{n}-t_{n-1}$. In case of reasonably small $R_{n}$ and bounded $\Psi_{n}$, the whole step-by-step procedure can be presented in the following way:
\begin{equation}\label{ts3}
\Psi_{0}\rightarrow\big[\Psi_{1}(t_{1})+R_{1}\big]\rightarrow\big[\Psi_{2}(t_{2})+R_{1}+R_{2}\big]\rightarrow\,\ldots\,\rightarrow\bigg[\Psi_{N}(t_{N})+\sum\limits_{n=1}^{N}R_{n}\bigg]\;.
\end{equation}

Further, for the certain single-step propagators with the acceptable precisely calculated step errors it can be proved:
\begin{equation}\label{ts4}
\lim\limits_{N\rightarrow\infty}\bigg[\Psi_{N}(t_{N})+\sum\limits_{n=1}^{N}R_{n}\bigg]=\Psi(T)\;.
\end{equation}

The derivation of various single-step propagators for various equations is a well known and widely discussed problem. For the time-depended Schr\"odinger equation the arbitray-order single-step propagators have been obtained and discussed in \cite{SSP}. For the Eq.(\ref{i1}) we use here equidistant time slicing with $t=t_{n}-t_{n-1}=T/N$ and a first-order single-step propagator in the form:
\begin{equation}\label{ts5}
\hat{P}_{n}=\left[1+\frac{\mathrm{i}\,t}{2}\,\frac{\partial^{2}}{\partial{}x^{2}}\right]\cdot{}\exp\bigg[-\mathrm{i}t\,U(x,n{}t)\bigg]+O(t^{2}).
\end{equation}
For the equidistant time-slicing we have, the error at each propagation step $t$ is proportional to $\sim{}O(t^{p+1})$, where $p\in\mathbb{N}$ is an order of the corresponding single-step propagator. The total error is $\sum\limits_{n=1}^{N}R_{n}\sim{}N\cdot{}t^{p+1}=T^{p+1}N^{-p}$. Therefore, to fulfill the fundamental limit Eq.(\ref{ts4}), it is sufficient to have any $p\geq{}1$. We note, that the form Eq.(\ref{ts5}) is not the one possible option, further other different single-step propagators (of arbitrary order and dimensionality) can be used within our general approach. More details on this point can be found in \cite{SSP}.

\section*{Translation representation of differential\\ operators and single-step propagators}
The group of translation operators $\hat{T}$ is Abelian group:
\begin{equation}\label{it1}
\begin{aligned}
&\hat{T}_{a}\hat{T}_{b}=\hat{T}_{b}\hat{T}_{a}=\hat{T}_{(a+b)}\,,\;\;\;
&\hat{T}_{0}=\mathrm{I}\,\,\,;
\end{aligned}
\end{equation}
where $\mathrm{I}$ is an identity element. Each translation operator ${\hat{T}_{a}}=\hat{T}(a)$ depends on parameter $a\in\mathbb{R}$ (or possibly $a\in\mathbb{C}$) which is usually called shift. We consider that the translation operators map functions within some certain space of functions relevant to the chosen problem statement. The domain of the functions can be of arbitrary dimensionality, we denote the related arguments as $x$ or ${\bm x}=(x_{1},\,...\,,x_{K})$, ($K\in{\mathbb{N}}$). For example, for a function $g(x)$ the translation operator $\hat{T}_{a}$ acts as follows: $\hat{T}_{a}g(x)=g(x+a)$.

We work mostly with smooth functions (from $\mathbb{C}^{\infty}$) of complex variables unless specified. The notations for the differential operators are usual:
\begin{equation}\label{it2}
\hat{D}^{m}_{\lambda}=\left(\lambda\,\frac{\partial}{\partial{x}}\right)^{m}=\lambda^{m}\frac{\partial^{m}}{\partial{x}^{m}}\;,
\end{equation}
where $m\in{\mathbb{N}}$ is an arbitrary operator power, and $\lambda\in\mathbb{R}$ (or possibly $\lambda\in\mathbb{C}$) is an arbitrary $x$-independent parameter. We also assume in notations that $\hat{D}^{0}_{\lambda}\equiv{\mathrm{I}}$, $\hat{D}^{1}_{\lambda}\equiv{}\hat{D}_{\lambda}$, and ($\hat{D}^{m}_{0}\equiv{0}$, $m>0$). 

The expansion of a translation operator into a series of differential operators is related to the Taylor expansion and can be obtained directly:
\begin{equation}\label{it4}
\hat{T}_{a}=\sum_{k=0}^{\infty}\frac{1}{k!}\,\hat{D}_{a}^{k}=\exp\left[\hat{D}_{a}\right]\;.
\end{equation}

The inverse expansion of a differential operator $\hat{D}_{\lambda}$ in the series of translation operators up to a certain order of $\lambda$ can be written differently. Further we consider the real shift translations unless specified. From Eq.(\ref{it4}) we can obtain:
\begin{subequations}\label{it5}
\begin{align}
&\hat{D}_{\lambda}=\frac{1}{2}\hat{T}_{\lambda}-\frac{1}{2}\hat{T}_{\text{-}\lambda}+O(\lambda^{3})\;,\\
&\hat{D}^{2}_{\lambda}=-2+\hat{T}_{\lambda}+\hat{T}_{\text{-}\lambda}+O(\lambda^{4})\;.
\end{align}
\end{subequations}
Using these equalities we can rewrite the single-step propagators replacing the differential operators by the translation operators. For instance, for the Eq.(\ref{i1}) and its single-step propagator Eq.(\ref{ts5}) we can obtain:
\begin{equation}\label{it6}
\hat{P}_{n}=\bigg[(1-2\mathrm{i})+\mathrm{i}\hat{T}_{{+}}+\mathrm{i}\hat{T}_{{-}}\bigg]\cdot{}\exp\bigg[-\mathrm{i}t\,U(x,n{}t)\bigg]+O(t^{2}),
\end{equation}
where  $T_{\pm}=T(\pm\tau)=\exp(\pm\tau\frac{\partial}{\partial{}x})$, and $\tau=\sqrt{t/2}$.

The extension of Eq.(\ref{it6}) to a $K$-dimentional case can be done by introducing the vector ${\bm x}=(x_{1},\,...\,,x_{K})$ and replacing the $\frac{\partial^{2}}{\partial{}x^{2}}$ with $\Delta=\sum\limits_{i=1}^{K}\frac{\partial^{2}}{\partial{}x_{i}^{2}}$. This leads to the following:
\begin{equation}\label{it8}
\hat{P}_{n}=\bigg[(1-2\mathrm{i}K)+\mathrm{i}\sum\limits_{i=1}^{K}\left(\hat{T}_{(+i)}+\hat{T}_{(-i)}\right)\bigg]\cdot{}\exp\bigg[-\mathrm{i}t\,U({\bm x},nt)\bigg]+O(t^{2}),
\end{equation}
where $\hat{T}_{(\pm{}i)}=\exp(\pm\tau\frac{\partial}{\partial{}x_{i}})$. The former is only one possible option for the $K$-dimentional single-step propagator of the first order.

For the reasons of more simple explanation of our methods we consider also another parabolic partial differential equation, namely a special one-dimensional diffusion equation with an arbitrary given potential $V(x,t)$. The significant difference form the Eq.(\ref{i1}) is that all the coefficients are real-valued:
\begin{align}\label{it9}
\frac{\partial}{\partial{}t}\,&\Phi=\left[\frac{1}{4}\,\frac{\partial^{2}}{\partial{}x^{2}}\,+V(x,t)\right]\Phi\,,\\
&\Phi(x,t)|_{t=0}=\Phi_{0}(x)\,.\nonumber
\end{align}
Therefore the single-step propagator to find the solution $\Phi(x,T)$ at an arbitary time $T$ can be written in a more simple form:

\begin{equation}\label{it7}
\hat{P}_{n}=\frac{1}{2}\bigg[\hat{T}_{{+}}+\hat{T}_{{-}}\bigg]\cdot{}\exp\bigg[t\,V(x,nt)\bigg]+O(t^{2})\;,
\end{equation}
where $t=T/N$, and $\hat{T}_{\pm}=\hat{T}(\pm\tau)=\exp(\pm\tau\frac{\partial}{\partial{}x})$ with $\tau=\sqrt{t/2}$ as stated above.

\newpage
\section*{Translation Evolution Grid}

In this section we present and develop a Translational Evolution Grid (TEG) representation as a necessary step towards the derivation of the compact-form solutions. TEG representation provides a mathematical description of the discrete solution to the partial differential equation via sum of the corresponding TEG paths. For simplicity we start from the consideration of the special diffusion equation Eq.(\ref{it9}) and the corresponding TEG of the variety-order 2.

We consider a discrete solution of Eq.(\ref{it9}) consisting of the product\footnote[1]{We use right-to-left order in the product of operators:
\begin{equation}\label{ig1}
\left\{\prod\limits_{n=1}^{N}\hat{P}_{n}\right\}\Phi=\left(\hat{P}_{N}\ldots\left(\hat{P}_{2}\left(\hat{P}_{1}\Phi\right)\right)\ldots\right)\;.\nonumber
\end{equation}
} of the corresponding single-step propagators Eq.(\ref{it7}):
\begin{equation}\label{ig1}
\Phi_{N}(x,T)=\frac{1}{2^{N}}\left\{\prod\limits_{n=1}^{N}\bigg[\big(\hat{T}_{+}+\hat{T}_{-}\big)\cdot{}e^{t\,V(x,nt)}\bigg]\right\}\Phi_{0}(x)\;,
\end{equation}
where $t=T/N$, $N$ is a time-slicing number, and $\lim\limits_{N\rightarrow\infty}\big[\Phi_{N}(x,T)\big]=\Phi(x,T)$.

The expression Eq.(\ref{ig1}) consists of $2^{N}$ terms, each includes some ordered combination of $\hat{T}_{+}$, $\hat{T}_{-}$, and  $e^{t\,V(x,nt)}$ operators. We can group them taking into account the resulting shift of the initial state $\Phi_{0}(x)$:

\begin{equation}\label{ig2}
\Phi_{N}(x,T)=\frac{1}{2^{N}}\sum\limits_{p=0}^{2^{N}-1}E_{p}\cdot\Phi_{0}(x+S_{p}\cdot\tau)=\sum\limits_{s=-N}^{N}{}W_{s}\cdot\Phi_{0}(x+s\cdot\tau)\;,
\end{equation}
where $E_{p}$ is a product of potential-shifted exponents which originated from the translated $e^{t\,V(x,nt)}$ operators, $S_{p}$ is a corresponding shift of the initial state, $W_{s}$ consists of a sum of $2^{-N}E_{p}$ terms with identical $S_{p}=s$. Following the Eq.(\ref{ig1}) and Eq.(\ref{ig2}) we can represent each of this normalized exponent product as a path on some grid which node positions depend on $n$ (discrete-time-related number) and $s$ (translation-related number), see Fig.1. We call such grid: Translation Evolution Grid (TEG), and each of the normalized potential-shifted exponent product: TEG-path. In Eq.(\ref{ig2}) each TEG-path with number $p$ has an expression: $2^{-N}E_{p}$. For definiteness, we choose also a direction of TEG-path: each path starts from origin ($n=0$ and $s=0$), includes ordered potential-shifted exponents product related to the grid nodes (${1}\leq{}n\leq{N}$), and ends at $n=N$ with some resulting shift $s_{r}=S_{p}$ as in the examples in Fig.1. 
Each intermediate point of a TEG-path on the grid (with some certain $n_{i}$ and $s_{i}$) corresponds to the term $\exp\left[t\,V(x+s_{i}\tau,T-(n_{i}-1)t)\right]$ in the product. Sometimes we will write for briefly $V_{k}(s_{k}\tau)$ meaning that this is a potential at the discrete-time $(N-k+1)t$ with a shift $s_{k}\tau$, for $1{}\leq{}k\leq{}N$. Thus, the $n$-direction is in some sense the opposite of the direction of evolution. The title "Translation Evolution Grid" comes from the interconnection of translation operators on a discrete grid and solutions that describe the evolution in the considered system. 

\clearpage
\begin{SCfigure}
  \includegraphics[scale=0.41]{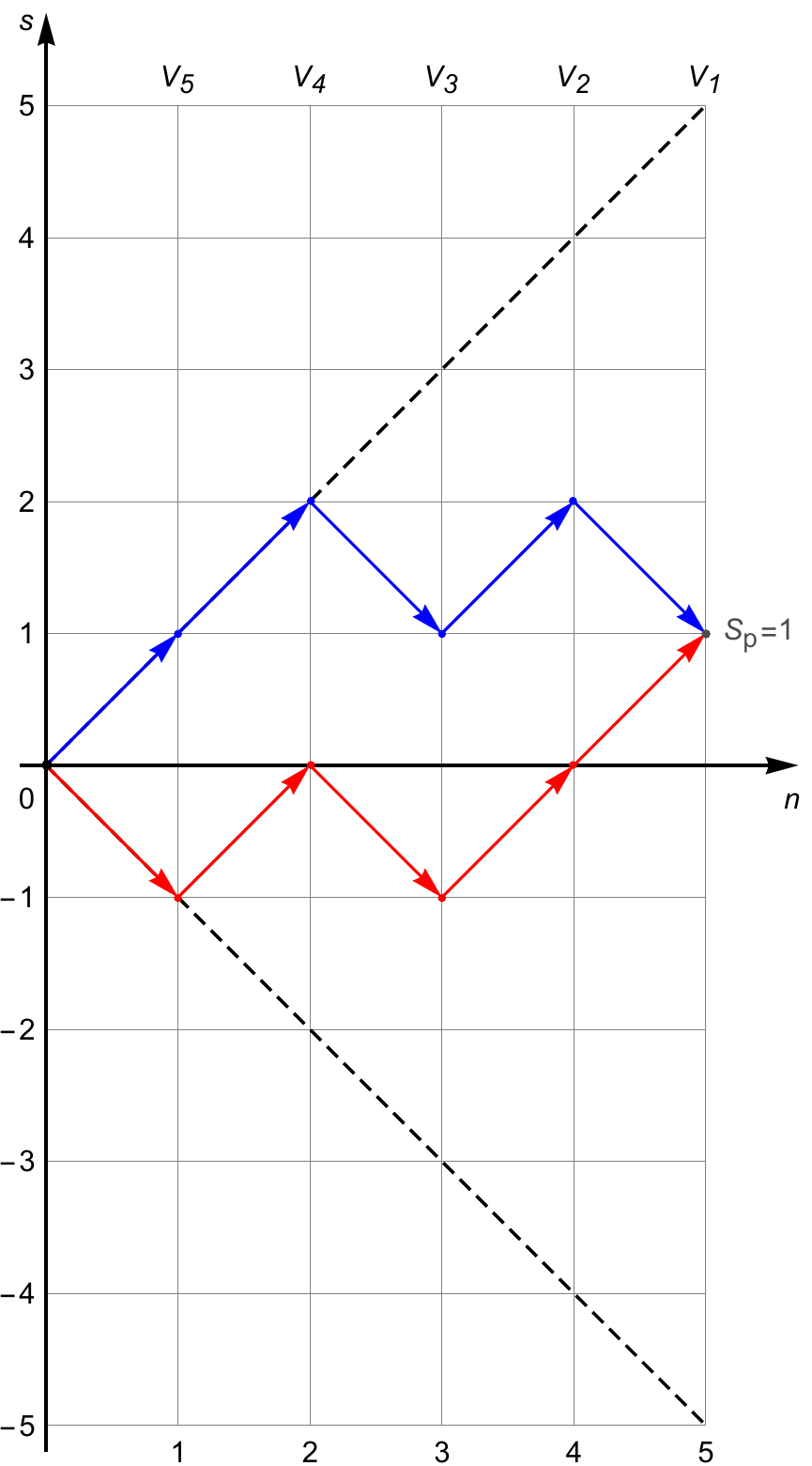}
\caption{A typical picture of the Translation Evolution Grid for the discretized equation Eq.(\ref{it9}) with two shown TEG-paths: \textcolor{blue}{upper} and \textcolor{red}{lower}. For the presented TEG the time-slicing number $N=5$, thus there are $2^{N}=32$ different TEG-paths. Each TEG-path can be encoded by the sequence of binary numbers: for example, each up-right move is $\{1\}$ and each down-right move is $\{0\}$. In this case the shown \textcolor{blue}{upper}-path is encoded by $\{\textcolor{blue}{11010}\}$, and the \textcolor{red}{lower}-path by $\{\textcolor{red}{01011}\}$. The related shifted-potential sum in the exponent product for the \textcolor{blue}{upper} TEG-path is: $V_{5}(\tau)+V_{4}(2\tau)+V_{3}(\tau)+V_{2}(2\tau)+V_{1}(\tau)$, and for the \textcolor{red}{lower} TEG-path is: \newline $V_{5}(-\tau)+V_{4}(0)+V_{3}(-\tau)+V_{2}(0)+V_{1}(\tau)$.
\newline}  
\end{SCfigure}

The TEG which is demonstrated in Fig.1 has a variety-order 2 because we have only two different translation operators $\hat{T}_{+}$ and $\hat{T}_{-}$ at each propagator step numbered by $n$-index in Eq.(\ref{ig1}). In general, the variety-order of a TEG corresponds to a number of unique translation operators including $\hat{T}(0)$ in the corresponding single-step propagator.

Since each $V_{k}(s_{k}\tau)$ for a certain $k$ is unique in principle, one needs to take into account and sum all the TEG-paths to solve the problem. One possible way to enumerate them is to use a sequence of binary numbers encoding the order of translation operators, see examples in Fig.1. Later in the manuscript, we will return to a consideration of a generic method for obtaining the corresponding summed expressions in a compact-form. Here we note only that the total number of TEG-paths with a certain resulting shift $s$ in Eq.(\ref{ig2}) is\, $\binom{N}{m}=\frac{N!}{m!(N-m)!}$, where $m=\frac{N-s}{2}$ is a number of down-right moves in the considered TEG-paths.

It is reasonable, in our opinion, to consider here one special case when $V(x,nt)\equiv{}0$. Then, the expression Eq.(\ref{ig1}) reads:
\begin{equation}\label{ig3}
\Phi_{N}(x,T)=\frac{1}{2^{N}}\left[\hat{T}_{+}+\hat{T}_{-}\right]^{N}\Phi_{0}(x)\;.
\end{equation}
The Eq.(\ref{ig3}) is not a compact-form expression since $\hat{T}_{+}$ and $\hat{T}_{-}$ are not accepted operations according to the compact-form definition. However, using the basic properties of translation operators, namely: $\hat{T}_{+}\hat{T}_{-}=\hat{T}_{-}\hat{T}_{+}$ and $\hat{T}_{+}^{N}=\hat{T}(N\tau)$, and binomial expansion we can rewrite it as follows:
\begin{equation}\label{ig4}
\Phi_{N}(x,T)=\frac{1}{2^{N}}\sum\limits_{k=0}^{N}\binom{N}{k}\hat{T}_{+}^{N-k}\hat{T}_{-}^{k}\Phi_{0}(x)=\frac{1}{2^{N}}\sum\limits_{k=0}^{N}\frac{N!\Phi_{0}(x+N\tau-2k\tau)}{k!(N-k)!}.
\end{equation}
This is a (discrete) compact-form solution of Eq.(\ref{it9}) when $V(x,t)\equiv{}0$, the factorial is equal to the related product: $N!=\prod_{i=1}^{N}i$. We can also transform it to the continuous compact-form expression applying the approximation of the binomial coefficients at $N\rightarrow\infty$ using similar considerations that were first presented in a pamphlet by Abraham de Moivre, see \cite{deMoivre}. A general method for obtaining analogous approximations and continuous expressions is discussed in the next section. 
In our case this gives:
\begin{equation}\label{ig5}
\binom{N}{k}\;\underset{(N\rightarrow\infty)}{\xrightarrow{\hspace*{1.3cm}}}\;\;\;\frac{2^{N}}{\sqrt{\pi{}N/2}}\cdot\exp\bigg[-\frac{2}{N}\left(k-\frac{N}{2}\right)^{2}\bigg]\;\;.
\end{equation}
Now using a new variable $y=\frac{1}{\sqrt{N/2}}\cdot\left(k-\frac{N}{2}\right)$ as $N$ tends to infinity $(N\rightarrow\infty)$ we can write an integral instead of sum in Eq.(\ref{ig4}) and obtain:
\begin{equation}\label{ig6}
\Phi(x,T)\;=\;\frac{1}{\sqrt{\pi}}\;\int\limits_{-\infty}^{+\infty}e^{-y^{2}}\Phi_{0}\left(x+y\sqrt{T}\right)dy\;.
\end{equation}
This is an exact compact-form solution of Eq.(\ref{it9}) when $V(x,t)\equiv{}0$, which can also be checked by direct substitution.

In the same way one can consider a TEG for the time-dependent Schr\"odinger equation Eq.(\ref{i1}) with the single-step propagator Eq.(\ref{it6}). In this case, except the analogous up-right moves and the down-right moves we have also horizontal moves to the right, corresponding to $(1-2\mathrm{i})$ term in Eq.(\ref{it6}). The discussed TEG has a variety order 3, thus a possible option to encode the TEG-paths is using a ternary (base 3) numeral system. The multi-dimensional TDSE of dimensionality $K$ with the single-step propagator Eq.(\ref{it8}) can be described by a TEG with the variety order (2$K$+1). The generalization is straightforward. 

We note here, that following the presented approach one can directly obtain TEG-path examples (in closed-form as normalized exponent of shifted-potential sum) related to a TEG of an arbitrary variety order. Further, one can sum a number of TEG-paths with some certain statistical/problem-dependent coefficients to obtain the approximation of the solution. For reasonable accuracy, the number of summed weighted examples of TEG-paths can sometimes be much less than the total number of TEG-paths. This can be pivotal for the study of complex multi-dimensional systems.

A general method to take into account and sum all the TEG-paths in a compact-form is presented in the next section.

\section*{Combination-Selection Integration}

In this section we consider a method to analyze the combinatorial expressions through the special integration technique. We first introduce an elementary function of two variables $m$ and $k$ ($m\in\mathbb{Z}$ and $k\in\mathbb{Z}$) which is written as a single integral:

\begin{equation}\label{ic4}
G_{I}(m,k)=\frac{1}{2\pi}\int\limits_{-\pi}^{\pi}e^{\mathrm{i}m\phi}e^{-\mathrm{i}k\phi}d\phi\,=\begin{dcases}
    1,\;\; & \text{if }\; k=m\\
    0,\;\; & \text{otherwise}
\end{dcases}
\end{equation}

This equality introduces an elementary indicator function through a simplest finite integral. Now we fundamentally expand this description to control the selection of some elements within the inner expressions. In that way we can represent some choice/logic operations under some objects using a finite number of integrals.

We introduce a finite set of some objects $\{\mathrm{\bf P}\}$ wich can be functions or operators. We introduce also a finite set of integer numbers $\{\mathrm{\bf m}\}$ associated to the structure, a finite set of integer numbers $\{\mathrm{\bf k}\}$ associated to the condition(s), and a finite set of integration variables $\{\mathrm{\bf Y}\}$ such that the elementary integration volume $d{\bf Y}=d\phi_{1}\cdot{}d\phi_{2}\,\cdots$. Then we can distinguish two finite sets of exponents $\{e^{\mathrm{i}\mathrm{\bf m}\mathrm{\bf Y}}\}$ and $\{e^{-\mathrm{i}\mathrm{\bf k}\mathrm{\bf Y}}\}$ where each element is an exponent of $(\pm\mathrm{i})$ product of some element from ($\{\mathrm{\bf m}\}$ or $\{\mathrm{\bf k}\}$) and some element form $\{\mathrm{\bf Y}\}$. Now we can write a general integral with an optional normalization constant $f$ in the following form:
\begin{equation}\label{ic5}
{\bm G}_{CS}(\mathrm{\bf P},\mathrm{\bf m},\mathrm{\bf k})=f\int\limits_{-\pi}^{\pi}\cdots\int\limits_{-\pi}^{\pi}\underbrace{S(\mathrm{\bf P}\,,\,e^{\mathrm{i}\mathrm{\bf m}\mathrm{\bf Y}})}_{\text{Structure}}\,\cdot{}\,\underbrace{C(e^{-\mathrm{i}\mathrm{\bf k}\mathrm{\bf Y}})}_{\text{Conditions}}\,d{\bf Y}\,.
\end{equation}

In this form $S(\mathrm{\bf P}\,,\,e^{\mathrm{i}\mathrm{\bf m}\mathrm{\bf Y}})$ and $C(e^{-\mathrm{i}\mathrm{\bf k}\mathrm{\bf Y}})$ provide some certain expressions with required combination-selection properties. This representation is called Combination-Selection Integration (CSI). Next we will specify the possible options by looking at some examples and solving some problems related to the main topic. 

Our first example is SCI for the specific combination. Assume we have a set of elements $\mathrm{\bf P}=\{P_{1},P_{2},P_{3}\}$ where $P_{i}P_{j}=P_{j}P_{i}$ and we need to write a sum of all products of two different elements from $\mathrm{\bf P}$, then one possible option is:
\begin{equation}\label{ic6}
\frac{1}{2\pi}\int\limits_{-\pi}^{\pi}(1+P_{1}e^{\mathrm{i}\phi})(1+P_{2}e^{\mathrm{i}\phi})(1+P_{3}e^{\mathrm{i}\phi})e^{-2\mathrm{i}\phi}d\phi\,=\\ P_{1}P_{2}+P_{2}P_{3}+P_{1}P_{3}\,.
\end{equation}

In case we have $\mathrm{\bf P}=\{\hat{P}_{1},\hat{P}_{2},\hat{P}_{3}\}$ where $\hat{P}_{i}\hat{P}_{j}\neq{}\hat{P}_{j}\hat{P}_{i}$ the corresponding expression for the pairs of different objects can be written as:
\begin{align}\label{ic7}
\frac{1}{2\pi}\int\limits_{-\pi}^{\pi}(\hat{P}_{1}e^{\mathrm{i}\phi}+\hat{P}_{2}e^{4\mathrm{i}\phi}+\hat{P}_{3}e^{9\mathrm{i}\phi})^{2}\cdot\left(e^{-5\mathrm{i}\phi}+e^{-10\mathrm{i}\phi}+e^{-13\mathrm{i}\phi}\right)d\phi\,= \nonumber \\
\hat{P}_{1}\hat{P}_{2}+\hat{P}_{2}\hat{P}_{1}+\hat{P}_{2}\hat{P}_{3}+\hat{P}_{3}\hat{P}_{2}+\hat{P}_{1}\hat{P}_{3}+\hat{P}_{3}\hat{P}_{1}\,.
\end{align}

Using the same approach we can obtain compact-form expressions for the various combinations defined by specific conditions. For instance, assume we have a set of elements $\mathrm{\bf P}=\{P_{1},\,\ldots\,,P_{N}\}$, with $P_{i}P_{j}=P_{j}P_{i}$ and we need to write a sum of all products of minimum $L$ and maximum $M$ different elements from $\mathrm{\bf P}$, then the compact-form expression $G_{S}$ is:
\begin{equation}\label{ic7}
G_{S}=\frac{1}{2\pi}\int\limits_{-\pi}^{\pi}\,\left[\prod\limits_{n=1}^{N}(1+P_{n}e^{\mathrm{i}\phi})\right]\,\cdot{}\left(\,\sum\limits_{k=L}^{M}e^{-\mathrm{i}k\phi}\,\right)d\phi\,.
\end{equation}

The proposed approach provides a mathematical tool to construct, analyze, and count some object configurations and countable structures. 
It can also be used to describe some combinatorial expressions. The properties of the binomial coefficient can be analyzed using its natural CSI representation:
\begin{equation}\label{ic8}
\binom{N}{k}=\frac{N!}{k!(N-k)!}=\frac{1}{2\pi}\int\limits_{-\pi}^{\pi}\,\left(1+e^{\mathrm{i}\phi}\right)^{N}\cdot{}e^{-\mathrm{i}k\phi}\,d\phi\,.
\end{equation}

CSI allows also to obtain various approximations. 
As an example, we derive Stirling-Moivre approximation of a factorial and discuss the calculation of its further corrections. By using exponent series definition and Eq.(\ref{ic4}) we can write for $\forall{B}$ the following:
\begin{equation}\label{ic9}
\frac{B^{N}}{N!}=\frac{1}{2\pi}\int\limits_{-\pi}^{\pi}\,\exp\left(B\,e^{\mathrm{i}\phi}\right)\,e^{-\mathrm{i}N\phi}\,d\phi\,.
\end{equation}

Now we choose $B=N$ and change the integration variable to $\varphi=\phi\sqrt{N}$, this gives the following:
\begin{equation}\label{ic10}
\frac{N^{N}}{N!}=\frac{1}{2\pi{}\sqrt{N}}\int\limits_{-\pi{}\sqrt{N}}^{\pi{}\sqrt{N}}\,\exp\left(N\,e^{\mathrm{i}\varphi/\sqrt{N}}\right)\,e^{-\mathrm{i}\varphi\sqrt{N}}\,d\varphi\,.
\end{equation}

Next, we use a series expansion: $\left(N\,e^{\mathrm{i}\varphi/\sqrt{N}}\right)=N+\mathrm{i}\varphi\sqrt{N}-\frac{\varphi^{2}}{2}+\sum\limits_{n=3}^{\infty}R_{n}$, where $R_{n}=\frac{(\mathrm{i}\varphi)^{n}}{n!(\sqrt{N})^{n-2}}$, and finally obtain the following:
\begin{equation}\label{ic11}
\frac{N^{N}}{N!}=\frac{e^{N}}{2\pi{}\sqrt{N}}\int\limits_{-\pi{}\sqrt{N}}^{\pi{}\sqrt{N}}\,\exp\left(-\frac{\varphi^{2}}{2}+\sum\limits_{n=3}^{\infty}R_{n}\right)\,d\varphi\;\underset{(N\rightarrow\infty)}\approx\;\frac{e^{N}}{2\pi{}\sqrt{N}}\cdot{}\sqrt{2\pi}\,.
\end{equation}

This is equivalent to the following approximation: $N!\approx{}\sqrt{2\pi{}N}\,\cdot{}N^{N}e^{-N}$ at $N\rightarrow\infty$. Additional corrections can be found by considering more accurate approximation (at $N\rightarrow\infty$) of integrals Eq.(\ref{ic10}-\ref{ic11}). Other various approximations, as well as exact equalities, can be derived in a similar way.

Several problems related to the topic of this manuscript can be solved by using CSI method. One key procedure for obtaining a compact-form solution of Eq.(\ref{i01}) is presented in the next section.

\newpage

\section*{Arbitrary base representation of a number:\\compact-form expressions for the coefficients}

In this section we apply CSI method to obtain compact-form expessions to the coefficients of the arbitrary base representation of an arbitrary given number. We consider the following base-representation theorem:

\underline{\textit{Theorem}}: For any given number $M\in {\mathbb{N}}$ and any $b\geq{}2$, ($b\in{\mathbb{N}}$ is the representation \textit{base}) there is a unique representation:
\begin{equation}\label{ib1}
M=a_{0}+a_{1}b^{1}+a_{2}b^{2}+\ldots +a_{l}b^{l} \,,
\end{equation}
with a certain $l\in{\mathbb{N}}$ such that $M<b^{l+1}$ (we call $l$ the representation length), and certain $0\leq{}a_{m}<b$ for $\forall{m}\in\{0,...,l\}$ (we call $a_{m}=a_{m}(b,M)\in{\mathbb{N}_{0}}$ the coefficients of the base-$b$ representation of a number $M$). 

\underline{Remark}: The existence of the representation Eq.(\ref{ib1}) can be proved by using Euclidean division or simply by induction, and uniqueness by contradiction, one possible variant can be found in \cite{Moll}. The uniquiness of the representation guaranties that any $b^{n}$ can not be presented as a sum of different degrees of $b$, which is useful for selection of some specified TEG-paths.
 
\underline{\textit{Proof of the base-representation theorem}}: We prove existence by induction on $M$. While for $M=1$ is clear, there are only two possible variants for $(M+1)$ assuming that for $M$ is true: 1) $a_{m}(b,M)=(b-1)$, $\forall{m}\in\{0,...,l\}$, then $M=\sum\limits_{m=0}^{l}(b-1)\cdot{}b^{m}=b^{l+1}-1$ and thus a new representation is: $(M+1)=b^{l+1}$. 2) there is a first number $\tilde{m}$ from the beginning such that $a_{\tilde{m}}(b,M)<(b-1)$, and we have\footnote[1]{If $\tilde{m}=0$, then according to the typical convention: $\sum\limits_{m=0}^{\tilde{m}-1}(b-1)\cdot{}b^{m}\equiv{}0$.}: $M=\sum\limits_{m=0}^{\tilde{m}-1}(b-1)\cdot{}b^{m}+a_{\tilde{m}}b^{\tilde{m}}+\sum\limits_{m=\tilde{m}+1}^{l}a_{m}(b,M)\cdot{}b^{m}$. Thus, a new representation is following: $(M+1)=(a_{\tilde{m}}+1)\cdot{}b^{\tilde{m}}+\sum\limits_{m=\tilde{m}+1}^{l}a_{m}(b,M)\cdot{}b^{m}$, with new coefficients are: $a_{m}(b,M+1)=0$ for $m<\tilde{m}$, $a_{\tilde{m}}(b,M+1)=\left[a_{\tilde{m}}(b,M)+1\right]$ for $m=\tilde{m}$, and $a_{\tilde{m}}(b,M+1)=a_{\tilde{m}}(b,M)$ for $\tilde{m}<m\leq{}l$.

The proof of uniqueness is by contradiction. Assume we have two different representations for one number with the first (index $\bar{m}$) two different coefficients $a'_{\bar{m}}\neq{}a''_{\bar{m}}$, subtracting these representations and then dividing by $b^{\bar{m}}$ we obtain: $(a'_{\bar{m}}-a''_{\bar{m}})+\sum\limits_{m=\bar{m}+1}^{l}(a'_{m}-a''_{m})\cdot{}b^{m-\bar{m}}=0$. The last term (a sum over $m$) is either zero or divisible by $b$ without a remainder, so the first term $(a'_{\bar{m}}-a''_{\bar{m}})$ is either zero or also divisible by $b$ without a remainder, which is a contradiction. 
\qed

In practice, finding the representation coefficients for a given $M$ is usually related to the sequential Euclidean division. In this way $M$ can be written in terms of unique quotient $q$ and remainder $r$ as follows:

\begin{equation}\label{ib2}
M=q\cdot{}b^{l}+r\,.
\end{equation}

From this equality, one can find that $a_{l}(b,M)=q$ and repeat the procedure in the same way to obtain $a_{l-1}(b,M)$ from the reminder $r$ being a new number for the next Euclidean division. From this follows that for the finding of a certain $a_{m}(b,M)$, one needs to perform $(l-m+1)$ operations of Euclidean division and these operations are recursive: any next step requires data from the previous one. Thus, the expression obtained in this way for some $a_{m}(b,M)$ is not a compact-form expression. 

In this section, we use the CSI method to obtain compact-form expressions for any $a_{m}(b,M)$ directly, without knowing a priori any other parameters/coefficients. In order to simplify the derivation we first consider a particular case: compact-form formulas to binary representation of an arbitrary given number $M$. Let us consider the following CSI:

\begin{equation}\label{ib3}
R_{\small \textcircled{\small \raisebox{-.9pt} {2}}}\left( M\right) =\dfrac{1}{2\pi }\int\limits ^{\pi }_{-\pi }\prod ^{N}_{n=0}\left[ 1+A_{n}\cdot{}e^{\mathrm{i}2^{n}\theta } \right]\cdot e^{-\mathrm{i}M\theta }d\theta \;,
\end{equation}
where $N>-1+\log_{2}M$ is equal to the representation length, and $A_{n}=A(n)$ is a set of some expressions that will be specified later. Since we have a unique binary representation of $M$ according to the base-representation theorem, this integral is equal to a unique product of $A_{i}$ with some certain set of $i$-indices. For example, for $M=13=1\cdot{}2^{0}+0\cdot{}2^{1}+1\cdot{}2^{2}+1\cdot{}2^{3}$ and any $N\geq{}3$ we have $R_{\small \textcircled{\small \raisebox{-.9pt} {2}}}\left( 13\right)=A_{0}\cdot{}A_{2}\cdot{}A_{3}$. In order to obtain the compact-form expression for the certain $a_{m}=a_{m}(2,M)$ we can choose $A_{n}=\left( 1+e^{\mathrm{i}2^{n}\varphi }\right)$ and apply additional integration over $\varphi$ with a CSI-condition term $e^{-\mathrm{i}2^{m}\varphi }$ (since $2^{m}$ is a unique base-$2$ representation of $2^{m}$). This gives the following compact-form expression:

\begin{equation}\label{ib4}
a_{m}(2,M) =\dfrac{1}{4\pi^{2} }\int\limits ^{\pi }_{-\pi }\int\limits ^{\pi }_{-\pi }\prod ^{N}_{n=0}\left[ 1+\left( 1+e^{\mathrm{i}2^{n}\varphi }\right)\cdot{}e^{\mathrm{i}2^{n}\theta }\right]\cdot e^{-\mathrm{i}2^{m}\varphi } e^{-\mathrm{i}M\theta }\;d\theta d\varphi \;.
\end{equation}

This procedure can be easily generalized to any arbitrary given representation base $b$. In the same way we consider:
\begin{equation}\label{ib5}
R_{\small \textcircled{\small \raisebox{-.9pt} {b}}}\left( M\right) =\dfrac{1}{2\pi }\int\limits ^{\pi }_{-\pi }\prod ^{N}_{n=0}\left[ 1+\sum\limits_{k=1}^{b-1} A_{n}(k)\, e^{\mathrm{i}k b^{n}\theta } \right]\cdot e^{-\mathrm{i}M\theta }d\theta \;,
\end{equation}
now we can choose the following variant $A_{n}(k)=\left( 1+k\cdot{}e^{\mathrm{i} 2^{n}\varphi }\right)$, and apply integral over $\varphi$ like before taking into account the base-representation uniqueness. The compact-form expression for $a_{m}(b,M)$ is as follows:

\begin{equation}\label{ib6}
\begin{aligned}
&a_{m}(b,M) =   \\ 
&\dfrac{1}{4\pi^{2} }\int\limits ^{\pi }_{-\pi }\int\limits ^{\pi }_{-\pi }\prod ^{N}_{n=0}\left[ 1+\sum\limits_{k=1}^{b-1} \left( 1+k\cdot{}e^{\mathrm{i} 2^{n}\varphi }\right)\,e^{\mathrm{i}k b^{n}\theta } \right]\cdot e^{-\mathrm{i}2^{m}\varphi } e^{-\mathrm{i}M\theta }\;d\theta d\varphi \;.
\end{aligned}
\end{equation}

The formula Eq.(\ref{ib6}) works as follows: for any given number $M$ and any given base $b$, if one takes any number $N>(\log_{b}M\,-1)$, Eq.(\ref{ib6}) gives the coefficients $a_{m}(b,M)$ of the unique base representation in the form of Eq.(\ref{ib1}). The given number $N$ is then equal to the representation length. Otherwise, if one takes such number $N$ that $M\geq{}b^{N+1}$ then the corresponding CSI Eq.(\ref{ib6}) gives: $a_{m}(b,M)\equiv{0}$ for $\forall{m}$. The universal general compact-form expression for the coefficients of the arbitrary base representation Eq.(\ref{ib6}) provides a way to obtain compact-form expressions for all the numbered TEG-paths.

\section*{Compact-form expression for each TEG-path}

Using the results of the previous sections, we can now derive compact-form expressions for each numbered TEG-path and then sum them to obtain a discrete compact-form solution of the corresponding differential equation. We start from the equation Eq.(\ref{it9}) and use the related single step propagator Eq.(\ref{it7}). As was shown in section \textit{Translation Evolution Grid}, the discrete solution of Eq.(\ref{it9}) consists of a sum of $2^{N}$ terms each corresponds to a certain TEG-path and a shifted initial state: 
\begin{equation}\label{cfteg2}
\Phi_{N}(x,T)=\sum\limits_{M=0}^{2^{N}-1}{P}_{M}\cdot\Phi_{0}(x+S_{M}\cdot\tau)\;,
\end{equation}
where ${P}_{M}=2^{-N}\cdot{}E_{M}$ represents a unique $M$-numbered TEG-path, and $S_{M}$ is an accumulative shift of the related TEG-path with number $M$.

Now, we establish a one-to-one correspondence between each ${P}_{M}$ and binary representation of the $M$ from the $[0\,,\,2^{N}-1]$ interval. To do this we use the results from the previous section with a generalization of Eq.(\ref{ib3}) for the two different types of the product inner expressions:
\begin{equation}\label{cfteg3}
R^{g}_{\small \textcircled{\small \raisebox{-.9pt} {2}}}\left( M\right) =\dfrac{1}{2\pi }\int\limits ^{\pi }_{-\pi }\prod ^{N-1}_{n=0}\left[ B_{n}+A_{n}\cdot{}e^{\mathrm{i}2^{n}\theta } \right]\cdot e^{-\mathrm{i}M\theta }d\theta \;.
\end{equation}

In order to derive a compact-form expression for each TEG-path we need to control the presence of each of the inner expressions ($A_{n}$ or $B_{n}$) at the each stage $n$. This requires additional CSI integration over the other variable $\phi$ with the related CSI-structure exponents $e^{\mathrm{i}2^{n}\phi}$. In addition, we choose for instance that each $B_{n}$ is responsibe for the corresponding down-move and each $A_{n}$ is responsible for the corresponding up-move on the TEG. Thus, as a possible option, we add an additional parameter $\gamma$ to describe this process as follows:
\begin{equation}\label{cfteg4}
\begin{dcases}
    A_{n}=\left(1+e^{-\mathrm{i}\gamma}\cdot{}e^{\mathrm{i}2^{n}\phi}\right)\;\; & \longrightarrow\;\text{corresponds to up-move}\;\;\;\;\;\;\;{\rotatebox[origin=c]{45}{\(\Longrightarrow\)}}\;\\
    B_{n}=\left(1+e^{+\mathrm{i}\gamma}\cdot{}e^{\mathrm{i}2^{n}\phi}\right)\;\; & \longrightarrow\;\text{corresponds to down-move}\;\;{\rotatebox[origin=c]{-45}{\(\Longrightarrow\)}}\;
\end{dcases}
\end{equation}
From this, we introduce a TEG-selection function $G_{\small \textcircled{\small \raisebox{-.9pt} {2}}}\left( M , \phi , \gamma \right)=G_{\small \textcircled{\small \raisebox{-.9pt} {2}}}\left( M \right)$:
\begin{equation}\label{cfteg5}
G_{\small \textcircled{\small \raisebox{-.9pt} {2}}}\left( M \right) =\dfrac{1}{2\pi }\int\limits ^{\pi }_{-\pi }\prod ^{N-1}_{n=0}\left[ \left(1+e^{\mathrm{i}\gamma+\mathrm{i}2^{n}\phi}\right) + \left(1+e^{-\mathrm{i}\gamma+\mathrm{i}2^{n}\phi}\right)e^{\mathrm{i}2^{n}\theta } \right] e^{-\mathrm{i}M\theta }d\theta.
\end{equation}

This function is equal to a product of $\left(1+e^{\pm\mathrm{i}\gamma}\cdot{}e^{\mathrm{i}2^{n}\phi}\right)$ terms where each term with $(+\mathrm{i}\gamma)$ corresponds to the down-move on the TEG and each term with $(-\mathrm{i}\gamma)$ corresponds to the up-move on the TEG. Using the appropriate CSI-integration over $\phi$ one can access to the each $n$-dependent term related to the point on TEG, while using the suitable CSI-integration over $\gamma$ one can find the corresponding accumulative shift. Thus, to find this $\gamma$-dependent shift at the position $k$ from the start of TEG-path with number $M$, we introduce a $\Lambda_{k}(M,\gamma)$ function:

\begin{equation}\label{cfteg7}
\Lambda_{k}(M,\gamma)=\dfrac{1}{2\pi }\int\limits ^{\pi }_{-\pi } G_{\small \textcircled{\small \raisebox{-.9pt} {2}}}\left( M , \phi , \gamma \right) \cdot \exp\left[\sum\limits_{j=0}^{k}(-\mathrm{i}\,2^{j}\phi)\right]d\phi \;.
\end{equation} 

Further, we sum: $\sum\limits_{j=0}^{k}(-\mathrm{i}\,2^{j}\phi)=-\mathrm{i}(2^{k+1}-1)\phi$. Let us consider, for example, the TEG-paths on Fig.1. The upper-path number is $M_{u}=1+2+0+8+0=11$ and the lower-path is $M_{l}=0+2+0+8+16=26$. From Eq.(\ref{cfteg7}) we have:  
\begin{align}
\label{cfteg8}\nonumber
&\Lambda_{1}(11,\gamma)=e^{-\mathrm{i}\gamma}\;,&&\Lambda_{1}(26,\gamma)=e^{+\mathrm{i}\gamma}\;,\\ \nonumber
&\Lambda_{2}(11,\gamma)=e^{-\mathrm{i}\gamma}e^{-\mathrm{i}\gamma}\;,&&\Lambda_{2}(26,\gamma)=e^{+\mathrm{i}\gamma}e^{-\mathrm{i}\gamma}\;,\\ \nonumber
&\Lambda_{3}(11,\gamma)=e^{-\mathrm{i}\gamma}e^{-\mathrm{i}\gamma}e^{+\mathrm{i}\gamma}\;,&&\Lambda_{3}(26,\gamma)=e^{+\mathrm{i}\gamma}e^{-\mathrm{i}\gamma}e^{+\mathrm{i}\gamma}\;,\\ \nonumber
&\Lambda_{4}(11,\gamma)=e^{-\mathrm{i}\gamma}e^{-\mathrm{i}\gamma}e^{+\mathrm{i}\gamma}e^{-\mathrm{i}\gamma}\;,&&\Lambda_{4}(26,\gamma)=e^{+\mathrm{i}\gamma}e^{-\mathrm{i}\gamma}e^{+\mathrm{i}\gamma}e^{-\mathrm{i}\gamma}\;,\\ \nonumber
&\Lambda_{5}(11,\gamma)=e^{-\mathrm{i}\gamma}e^{-\mathrm{i}\gamma}e^{+\mathrm{i}\gamma}e^{-\mathrm{i}\gamma}e^{+\mathrm{i}\gamma}\;.&&\Lambda_{5}(26,\gamma)=e^{+\mathrm{i}\gamma}e^{-\mathrm{i}\gamma}e^{+\mathrm{i}\gamma}e^{-\mathrm{i}\gamma}e^{-\mathrm{i}\gamma}\;. \nonumber
\end{align}

Using the function $\Lambda_{k}(M,\gamma)$ one can obtain each shifted potential for the TEG-path with number $M$. The following CSI over $\gamma$ gives:
\begin{equation}\label{cfteg9}
V_{N-k}\bigg|_{M}=\dfrac{1}{2\pi }\int\limits ^{\pi }_{-\pi } \Lambda_{k}(M,\gamma)\,\cdot \left[\sum\limits_{s=-N}^{N} e^{\mathrm{i}s\gamma}\cdot{}V_{N-k}(x+s\tau)\right] d\gamma \;.
\end{equation} 

To obtain the product of the potential-shifted exponents $E_{M}$ we need to consider according Eq.(\ref{ig2}) the following expression:  
\begin{equation}\label{cfteg10}
E_{M} = \,\exp\left[ \dfrac{t}{2\pi } \sum\limits_{k=1}^{N}\sum\limits_{s=-N}^{N} \int\limits ^{\pi }_{-\pi } e^{\mathrm{i}s\gamma}\,\Lambda_{k}(M,\gamma)\cdot{}V_{N-k}(x+s\tau) d\gamma \right] \;.
\end{equation}

Now we can write a compact-form expression for the TEG-path of Eq.(\ref{it9}) $P_{M}=P_{M}(x,N,T)$ avoiding some extra notation and definitions. We take into account that $t=T/N$, $\tau=\sqrt{T/2N}$, $\sum\limits_{j=0}^{k}(-\mathrm{i}\,2^{j}\phi)=-\mathrm{i}(2^{k+1}-1)\phi$ and obtain:
\begin{equation}\label{cfteg11}
\begin{aligned}
&{P}_{M}=2^{-N}\,\times \\
&\exp\bigg[\sum\limits_{k=1}^{N}\sum\limits_{s=-N}^{N}\int\limits ^{\pi }_{-\pi }\int\limits ^{\pi }_{-\pi } G_{\small \textcircled{\small \raisebox{-.9pt} {2}}}\cdot V\left(x+s\tau,T-\frac{kT}{N}\right) \, \dfrac{T\,e^{\mathrm{i}s\gamma-\mathrm{i}(2^{k+1}-1)\phi}}{4\pi^{2}N } d\phi d\gamma \bigg].
\end{aligned}
\end{equation}

This general compact-form expression for the TEG-path of the number $M$ can further be rewritten in other forms, taking into account the expression for the TEG-selection function $G_{\small \textcircled{\small \raisebox{-.9pt} {2}}}=G_{\small \textcircled{\small \raisebox{-.9pt} {2}}}\left( M , \phi , \gamma \right)$ Eq.(\ref{cfteg5}) and, possibly, some properties of potential $V(x,t)$ in Eq.(\ref{it9}).

To extend this technique for the TEG-paths on TEG of arbitrary variety-order, we generalize the TEG-selection function Eq.(\ref{cfteg5}) for the arbitrary base-$b$ representation and a set of inner expressions $F_{k}(n)$ similar as in Eq.(\ref{ib5}).

\begin{equation}\label{cfteg12}
G_{\small \textcircled{\small \raisebox{-.9pt} {b}}}\left( M \right) =\dfrac{1}{2\pi }\int\limits ^{\pi }_{-\pi }\prod ^{N-1}_{n=0}\left[ F_{0}(n) + \sum\limits_{k=1}^{b-1} F_{k}(n)\, e^{\mathrm{i}k b^{n}\theta } \right] e^{-\mathrm{i}M\theta }d\theta,
\end{equation}
where $F_{m}(n)=\left[1+f_{m}(n,\gamma)\cdot{}e^{\mathrm{i}2^{n}\phi}\right]$ for $m\in{}[0,\,...\,,b-1]$ and function $f_{m}(n,\gamma)$ describes the corresponding $\gamma$-dependent operation (a certain shift for example) at the stage $n$. For instance, for the Eq.(\ref{i1}) with the corresponding single step propagator Eq.(\ref{it6}) and the base $b=3$, the functions are following: $f_{0}(n,\gamma)=1$ (horizontal move), $f_{1}(n,\gamma)=e^{-\mathrm{i}\gamma}$ (up-move), $f_{2}(n,\gamma)=e^{+\mathrm{i}\gamma}$ (down-move). For the multi-dimensional equation one can establish shift parameters for different dimensions by choosing functions $e^{\pm\mathrm{i}\gamma_{x}}$, $e^{\pm\mathrm{i}\gamma_{y}}$, $e^{\pm\mathrm{i}\gamma_{z}}$, etc. For the single step propagators of higher accuracy one can use also functions like $e^{\pm{2}\mathrm{i}\gamma}$, $e^{\pm{3}\mathrm{i}\gamma}$ or other specific combinations.

To obtain compact-form TEG-path expressions we need also to calculate the exponent normalizing factor $C_{M}$. In case of Eq.(\ref{it9}) we have a universal factor for all TEG-paths equal to $2^{-N}$. For the Eq.(\ref{i1}) and related Eq.(\ref{it6}) each up- or down-move gives a factor $(\mathrm{i})$ while each horizontal move a factor $(1-2\mathrm{i})$. Using the TEG-selection function in the form Eq.(\ref{cfteg12}) one can easily calculate the exponent prefactor $C_{M}$. However, in this case, we do not need to keep the choice of the factors at the intermediate stages (numbered with $n$-index), so according to the base representation theorem we can write a simpler CSI:
\begin{equation}\label{cfteg13}
C_{M} =\dfrac{1}{2\pi }\int\limits ^{\pi }_{-\pi }\prod ^{N-1}_{n=0}\left[ (1-2\mathrm{i}) + \mathrm{i} e^{\mathrm{i} 3^{n}\theta } + \mathrm{i} e^{2\mathrm{i} 3^{n}\theta } \right] e^{-\mathrm{i}M\theta }d\theta\,.
\end{equation}

The generalization to the multi-dimensional case is straightforward.
\newpage

Finally, to obtain a compact-form expression for the TEG-path $P_{M}$ in case of the TEG of the variety-order $(2K+1)$, like in Eq.(\ref{it8}), we need to use base $b=(2K+1)$ and write the following product:

\begin{equation}\label{cfteg14}
\resizebox{.9\linewidth}{!}{
\begin{math}
\begin{aligned}
&{P}_{M}=C_{M}\,\cdot\,E_{M}\;, \\
&C_{M}=\dfrac{1}{2\pi }\int\limits ^{\pi }_{-\pi }\prod ^{N-1}_{n=0}\left[ (1-2\mathrm{i}K) +  \sum\limits_{k=1}^{b-1} \mathrm{i}e^{\mathrm{i}k b^{n}\theta } \right] e^{-\mathrm{i}M\theta }d\theta\,,\\
&E_{M}=\\
&\exp\bigg[\sum\limits_{\kappa=1}^{N}\sum\limits_{{\bf s}}\oint\limits ^{\pi }_{-\pi }\int\limits^{\pi }_{-\pi} G_{\small \textcircled{\small \raisebox{-.9pt} {b}}}\cdot U\left({\bf x}+{\bf s}\tau,T-\frac{\kappa T}{N}\right)  \dfrac{-\mathrm{i}T\,e^{\mathrm{i}{\bf s}\pmb{\gamma}-\mathrm{i}(2^{\kappa+1}-1)\phi}}{(2\pi)^{K+1}N } d\phi d\pmb{\gamma} \bigg].
\end{aligned}
\end{math}
}
\end{equation}

In this expression ${\bf x}=(x_{1},\,...\,,x_{K})$ is a vector, ${\bf s}=(s_{1},\,...\,,s_{K})$ represents shifts in different coordinates such that the corresponding sum over ${\bf s}$ includes the sum over each coordinate ${ s_{i}}$ form ${ (-Ns_{i})}$ to ${ (+Ns_{i})}$, $\pmb{\gamma}$ is a related CSI condition vector such that ${\bf s}\pmb{\gamma}=(s_{1}\gamma_{1},\,...\,,s_{K}\gamma_{K})$ with $d\pmb{\gamma}=d\gamma_{1}\cdots{}d\gamma_{K}$, and $G_{\small \textcircled{\small \raisebox{-.9pt} {b}}}$ is a TEG-path selection function Eq.(\ref{cfteg12}) with the corresponding functions $f_{\kappa}(n,\pmb{\gamma})=\exp[\pm{}\gamma_{i}]$ describe the shifts on the considered TEG.

\section*{Compact-form solutions to the considered\\ partial differential equations}

In this section we summarize the results from the entire manuscript and write discrete compact-form solutions to the special diffusion equation Eq.(\ref{it9}) and the time-dependent Schr\"odinger equation Eq.(\ref{i1}) in one dimension with its multi-dimensional extension. We also discuss possible further modifications and simplifications with some options to obtain continuous compact-form expressions.

We start with a discrete compact-form solution to the special diffusion equation Eq.(\ref{it9}). According to the Eq.(\ref{ig2}) we need a compact form expression for the TEG-path $P_{M}$ and a compact-form expression for the coresponding accumulative shift $S_{M}$, both for arbitrary given number $M$. The compact-form expression for $P_{M}$ is presented in Eq.(\ref{cfteg11}), and for $S_{M}$ we apply the similar considerations as for the $C_{M}$ Eq.(\ref{cfteg13}) in the previous section. The compact-form expression is following:

\begin{equation}\label{ivcomp1}
S_{M} =\dfrac{1}{4\pi^{2} }\sum\limits_{k=-N}^{N}\int\limits ^{\pi }_{-\pi }\int\limits ^{\pi }_{-\pi }\prod ^{N-1}_{n=0}\left[ e^{+\mathrm{i}\gamma} + e^{-\mathrm{i}\gamma} e^{\mathrm{i} 2^{n}\theta }  \right] \cdot{}k\,e^{\mathrm{i}k\gamma}e^{-\mathrm{i}M\theta }d\theta{}d\gamma\,.
\end{equation}

By applying the CSI method we can write the solution of Eq.(\ref{it9}) in another compact-form by using the accumulative shift as a CSI parameter:

\begin{equation}\label{ivcomp2}
\begin{aligned}
&\Phi_{N}(x,T)=\sum\limits_{M=0}^{2^{N}-1}{P}_{M}\cdot\Phi_{0}(x+S_{M}\cdot\tau)=\;\\
&\sum\limits_{M=0}^{2^{N}-1}\sum\limits_{k=-N}^{N}\dfrac{P_{M}}{4\pi^{2}}\int\limits ^{\pi }_{-\pi }\int\limits ^{\pi }_{-\pi }\prod ^{N-1}_{n=0}\left[ e^{\mathrm{i}\gamma} + e^{-\mathrm{i}\gamma} e^{\mathrm{i} 2^{n}\theta }  \right] \cdot{}\Phi_{0}(x+k\tau)\,e^{\mathrm{i}(k\gamma-M\theta)}d\theta{}d\gamma,
\end{aligned}
\end{equation}
where $P_{M}$ is a compact-form expression for the TEG-path, see Eq.(\ref{cfteg11}), with a corresponding TEG-selection function $G_{\small \textcircled{\small \raisebox{-.9pt} {2}}}$, see Eq.(\ref{cfteg5}).

The generalization of Eq.(\ref{ivcomp2}) for the time-dependent Schr\"odinger equation can be implemented in the same way as discussed in the previous section for the generalization of TEG-path expressions. For the Eq.(\ref{i1}) the solution is presented in Eq.(\ref{i4}). A discrete compact-form solution for the multi-dimensional case:
\begin{equation}\label{ivcomp3}
\begin{aligned}
&\psi_{N}\left({\bf x},T\right) = \sum ^{(2K+1)^{N}-1}_{M=0}P_{M}\cdot \Psi _{0}\left( {\bf x}+{\bf S}_{M}\cdot \tau \right)=\sum ^{(2K+1)^{N}-1}_{M=0}\,\sum\limits_{{\bf s}}\,\dfrac{E_{M}}{(2\pi)^{K+1}}\times\\
&\oint\limits^{\pi }_{-\pi }\int\limits ^{\pi }_{-\pi }\prod ^{N-1}_{n=0}\left[ (1-2\mathrm{i}K) +  \sum\limits_{k=1}^{b-1} \mathrm{i}F_{k}(n,\pmb{\lambda})e^{\mathrm{i}k b^{n}\theta } \right]\cdot \Psi _{0}\left({\bf x}+{\bf s} \tau \right) e^{\mathrm{i}({\bf s}\pmb{\lambda}-M\theta)}d\theta{}d\pmb{\lambda}.
\end{aligned}
\end{equation}

In this equality the compact-form expression for the $P_{M}$ and its exponential part $E_{M}$ are presented in Eq.(\ref{cfteg14}) together with the explanation of a sum over ${\bf s}$ and an integral over $\pmb{\lambda}$, and for the functions $F_{k}(n,\pmb{\lambda})$ responsible for the shifts on the corresponding TEG one can see the Eq.(\ref{cfteg12}) and the related description.

The discrete compact-form solutions can be possibly transform to continuous compact-form expressions using various techniques. However, the utility of such transformations is questionable, in our opinion, since the discrete structure of the solutions makes naturally preferable the discrete options for the further analysis. Although the derivation of continuous compact-form expressions is not the purpose of the current manuscript, we discuss here some possible options. Since the time-slicing number $n$ is a discrete analog of time, it is natural to pass to the limit and transform the product into some integral, something like this: $\prod\limits^{N-1}_{n=0}\left(\cdots\right)\underset{(N\rightarrow\infty)}{\xrightarrow{\hspace*{1.3cm}}}\,\exp\Big[\int\,\log\left(\cdots\right)\,g(t)dt\Big]$ which can give a reasonable continuous representation of TEG-selection function and related expressions. Further, it would be preferable, in our opinion, to obtain the solution in the form structurally similar to Eq.(\ref{ig6}), namely:
\begin{equation}\label{ivcomp4}
\Phi_{N}(x,T)\underset{(N\rightarrow\infty)}{\xrightarrow{\hspace*{1.3cm}}}\,\int\limits_{-\infty}^{+\infty}\Omega\Big[V(x,t),y,T\Big]\cdot\Phi_{0}\left(x+y\sqrt{T}\right)dy\;,
\end{equation}
where $\Omega|_{T=0}=\Omega|_{V\equiv{}0}=\pi^{-\frac{1}{2}}e^{-y^{2}}$ is a functional depending on the potential and describing the density of the TEG-paths in the continuous representation.

We suggest, however, that a discrete calculus using multiple weight-modified TEG-paths with a proper distribution may be preferable for analysis.

\section*{Conclusions}

The general concept we develop in the manuscript can be formulated in terms of three ideas aimed at solving partial differential equations and related problems. The first one is a proper time discretization when using single-step propagators with a choosen overall accuracy. The second idea is an accuracy-justified replacement of all integrals and derivatives by a superposition of translation operators and potential functions. The third idea is the establishment of a one-to-one correspondence between the operator sequences and sets of natural numbers, and a proper account of all necessary for solving combinations. On this way, for the purposes of our work, we combine some methods of differential equations, enumerative combinatorics, and number theory. We believe that it will pave the way towards deeper understanding of considered systems.

\section*{Acknowledgements}

Author would like to thank his family for their various comprehensive support, and Arkady Gonoskov for the useful discussions.

\end{document}